# An ultra-cold, molecular Rydberg plasma with exceptionally long lifetime and strongly-coupled properties formed by threshold laser excitation in the expansion region of a supersonic jet

Jingwei Guo[1, 3], Mehrvash Varnasseri[1, 3], Mikko Riese[1, 2, 3] and Klaus Müller-Dethlefs[1, 3] *

[1] The Photon Science Institute, [2] Dalton Nuclear Institute, [3] School of Chemistry,
The University of Manchester, Alan Turing Building, Manchester M13 9PL, United Kingdom

An ultra-cold molecular plasma with extraordinarily long lifetime (~0.5 *ms*) is generated under strong collisional cooling conditions in the expansion region of a seeded supersonic jet expansion close to the nozzle. A resonant two-photon one-color laser process excites para-difluorobenzene molecules into the high-n Rydberg threshold region. Disorder heating during plasma formation is quenched by the high collision rate in the expansion, which keeps the ions at the translational jet temperature of 0.2*K*-0.7*K*. The Coulomb coupling parameter $\Gamma_i$ for the ions is expected to be *ca*. 230-820.

* corresponding author: k.muller-dethlefs@manchester

Plasmas are common in the universe. The "strength" of a plasma can be characterized by the Coulomb coupling parameter $\Gamma_\alpha$, where $\alpha=i$ stands for the ions and $\alpha=e$ for the electrons [1]:

$$\Gamma_\alpha = E_{interaction}/E_{thermal} = \frac{e^2}{4\pi\varepsilon_0 a} / k_B T_\alpha = 2.69 \times 10^{-5} \left[\frac{n}{10^{12} cm^{-3}}\right]^{1/3} \left[\frac{T_\alpha}{10^6 K}\right]^{-1}$$

$E_{interaction}$ is the Coulomb energy of the plasma, $E_{thermal}$ is the thermal energy of the ions (or electrons), $e$ is the electron charge, $a=(3/4\pi n)^{1/3}$ is the Wigner-Seitz radius and $n$ is the plasma number density. Plasmas are considered to be strongly-coupled when $\Gamma>1$. Current plasma theory predicts that for $1<\Gamma_i<170$ the plasma should show liquid-like short-range spatial correlations and solid-like long-range crystallization properties for $\Gamma_i>170$ [2,3].

The generation of a strongly-coupled plasma in the laboratory is a challenging problem. Equation (1) shows that a high density and/or a low temperature is required. A modern approach to reach strongly coupled atomic plasma conditions comes from experiments in magneto-optical traps. In these experiments a laser-induced plasma is formed in two steps [4,5]. First, the atoms are cooled to a few $\mu K$ and typically more than $10^6$ atoms are trapped in a $mm^3$. In the second step the atoms are resonantly laser-ionized or excited to high-*n* Rydberg states leading to plasma formation. Two recent reviews summarize the current state of the art both from the experimental and the theoretical point of view [5,6]. The typical densities ($10^9$ to $10^{10}/cm^3$) and temperatures obtained (10 $\mu K$) for the neutral atoms in an MOT should make the formation of a strongly-coupled plasma easily achievable. For instance, for an ionization (Rydberg formation) rate of 10%, a $\Gamma_i$ bigger than $10^5$ should be possible. However, what has been found is that the plasma formed from MOT cooled atoms and pulsed laser Rydberg excitation is far away from equilibrium due to the lack of strong correlation between the neutral atoms. Once the plasma is formed the local potential energy will be converted to kinetic energy of the ions and electrons, a process known as disorder heating [7]. Another important effect that leads to rapid heating of the electrons is three-body recombination. Hence on the $\mu s$ timescale the plasma will be heated up to several *K*, leading to rapid plasma expansion, reducing $\Gamma$ close to unity and maximum plasma lifetimes to *ca*. tens of $\mu s$ [5].

A novel way to produce an ultra-cold, or even strongly coupled plasma comes from the idea of forming a plasma from molecules seeded in a supersonic jet. Though the temperature $T_{trans}$ around a few tenths of *K* of the molecules is higher than $T_{initial}$ in a MOT, the collisions in the supersonic jet expansion prevent disorder heating and its associated rapid decrease of $\Gamma$. Another advantage is that this method can be applied to any substance, which can be brought into the gas phase, while plasmas formed in an MOT are restricted to atoms. The investigation of the influence of the internal degrees of freedom of a molecule on the formation of a plasma offers completely new perspectives.

Completely independent from our work, Grant and co-workers [8,9] have shown the formation of an ultra-cold plasma in a collisionless molecular beam of NO (skimmed supersonic jet at a distance of 10.5 *cm* from the nozzle) with an observation time of up to 30 $\mu s$. In their experiments the NO was excited by 1+1' resonant two-photon excitation into high-n Rydberg states, which evolved into a plasma *indirectly* by avalanche ionization.

Here we present a novel way of *directly* producing a plasma with strongly-coupled properties by resonant laser excitation of *para*-difluorobenzene (*p*DFB) to the Rydberg threshold region in the high-density, high-collision rate expansion region of a pulsed supersonic gas jet very close to the nozzle. We are particularly interested in finding out if this plasma can be formed instantaneously, *i.e.* not via the intermediate step of formation of strongly perturbed individual Rydberg states. Such a collective excitation to an electron gas, *i.e.* a plasma, should be inevitable when the individual Rydberg electron orbit radii corresponding to the excitation energy are much larger than the ion's Wigner-Seitz radius, a conjecture that we are going to substantiate.

Excitation into the high-n Rydberg region corresponding to $n \approx 200$ corresponds to a Rydberg electron orbit radius of around 2 $\mu m$. Compared to that, the Wigner-Seitz radius can be estimated from the molecular density $n$ at a distance $z$ from the nozzle $n(z)=I(0)_{id}/(u(z)*z^2)$, where $I(0)_{id}$ and $u(z)$ are the ideal centre-line intensity of a skimmerless source and the flow velocity, respectively [10,11]. Because $u(z)$ reaches its final value $u_\infty$ (0.6 $mm/\mu s$ for argon) after a few nozzle diameters, $u_\infty$ can be used instead. The centre-line intensity is given by $I(0)_{id}=\kappa(N'/\pi)$, $\kappa$ is the peaking factor (2.01 for argon), $N'=f(\gamma)n_0\alpha_0\pi R^2$ the flow through the nozzle, $f(\gamma)=0.531$ for argon, $n_0$ is the reservoir number density, $R_0$ is the nozzle radius and $\alpha_0=(2k_BT_0/m)^{1/2}$ is the most probable velocity in the reservoir of $T_0 =293K$ for molecular mass $m$. For a pure argon (1 *bar*, $n_0=2.4*10^{19}$ $cm^{-3}$) expansion the flow through the nozzle of 0.5 *mm* diameter is given by $N'=8.59*10^{20}$ $s^{-1}$, which gives a centre-line intensity of $I(0)_{id}=5.497*10^{20}$ *molecules*$*s^{-1}sr^{-1}$ and $n=2.247*10^{17}$ $cm^{-3}$ at a distance of $z=2$ *mm*. At room temperature *p*DFB has a vapor pressure of about 10 *mbar* [12], and with a reasonable ionization rate of 10% the estimated density of *p*DFB$^+$ ions is *ca.* $2.2*10^{14}$ $cm^{-3}$ giving a Wigner-Seitz radius of around 100 *nm*, which is much smaller than the Rydberg radius, thus justifying our assumption. For a translational jet temperature $T_{trans}$ of 0.2$K$ and 0.7$K$, as known from Refs 13, 14, 15, we find $\Gamma_i$ in the range from 820 to 230. Due to ionization in the high collision rate region, the ions formed are translationally cooled to $T_{trans}$ by collisions with the carrier gas argon atoms, and this process works against disorder induced heating of the ions.

Figure 1 shows a schematic diagram of the experimental set-up. The *p*DFB seeded in argon is introduced through a pulsed (General) valve with a 0.5 *mm* diameter nozzle into the first stage of the vacuum system (pumped by 2000 *l/s*). Close to the nozzle ($z = 0 – 15$ *mm*) the pDFB is excited into the high-n Rydberg region converging to a vibrational state of the ion, in the same way as in ZEKE and MATI spectroscopy [16], using the frequency-doubled output from a Radiant Narrowscan dye laser (range 261 to 270 *nm*, FWHM 0.05 $cm^{-1}$). The dye laser is calibrated against the iodine absorption spectrum. The UV pulse has a typical energy of 0.8 – 1 *mJ/pulse* with a Gaussian beam profile of 2 *mm* diameter. The second stage of the vacuum chamber is separated from the first stage by a 6 *mm diam.* skimmer and comprises an orthogonal acceleration time-of-flight mass spectrometer consisting of a dual stage acceleration region (apertures 1 to 3 in Fig. 1), ion optics (4-6), a drift region and a multi-channel plate detector (MCP) (pumped by 1000 *l/s*). Normally, high-n Rydberg excitation is performed by a two-color, two-photon process. For *p*DFB, however, it has been shown that Rydberg excitation is possible in a one-color two-photon process when the photon energy is accidentally resonant not only with a $S_1 \leftarrow S_0$ transition but also with high-n Rydberg $\leftarrow S_1$ transitions [17]. This makes the experimental realization of a plasma possible just with a single pulsed dye laser.

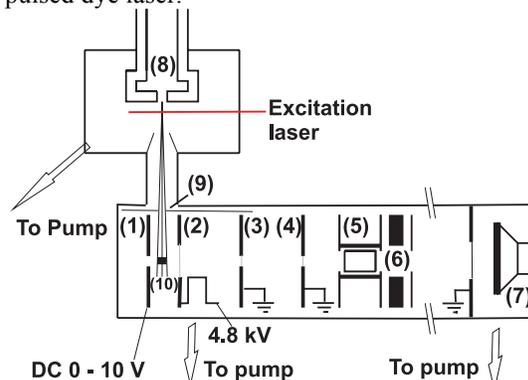

Figure 1: Schematic diagram of the experiment set-up. (1-4) apertures 1-4; (5) X -Y- deflectors; (6) ion lens; (7) MCP detector; (8) General valve; (9) 4 *mm* slit; (10) plasma cloud

One important criterion for the formation of a strongly-coupled plasma is that its lifetime is expected to be very long. To find out if a signal could be detected after ca. 0.5*ms*, *i.e.* the jet TOF from the nozzle to the centre line of the mass spectrometer, a first experiment was carried out by using Wiley-McLaren extraction [18] *i.e.* by simultaneously applying HV pulses of 1810 *V* and 1600 *V* on ap. 1 and 2, respectively, and recording the *p*DFB$^+$ ion TOF signal as function of laser wavelength. The HV pulses were applied ~ 500 $\mu s$ after the laser excitation (laser beam at $z=2mm$), which equals the arrival time of the plasma at the central line of the time-of-flight mass spectrometer ($v=0.64$ $mm/\mu s$, distance=320 *mm*). Figure 2 shows the obtained spectral scan in the region from 38000 to 38200 $cm^{-1}$. Under these conditions an extremely strong signal (0.5*V* in 0.2 $\mu s$ FWHM) is observed at 38090.4 $cm^{-1}$, corresponding to the transition into the $3^1$ vibrational state of the $S_1$ state. Further absorption of a second photon of the same energy leads to transitions into the high-n Rydberg state region converging to the $3^14^1$ and $3^15^130^1$ vibrational thresholds of the ion and this transition has been chosen to probe the plasma formation.

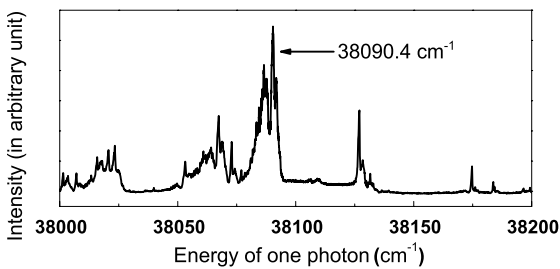

Figure 2: Laser wavenumber scan from 38000 to 38200 $cm^{-1}$

There are three species, which need to be considered: i) pDFB cations, ii) neutral pDFB Rydberg molecules and iii) the plasma. In order to find out what is present between ap. 1 and 2 ca. 500 µs after laser excitation, the effect of applying small DC voltages on ap. 1 (i.e. a small DC field between aps. 1 and 2) and a HV pulse of 4.8 kV on aperture 2 was studied. With this chosen detection scheme only charged species will be detected which i) are accelerated by the weak DC field, pass through the grid of ap. 2 and ii) are accelerated by the HV pulse on ap. 2 (800 V/cm between aps.2 and 3). The resulting measured time-of-flight spectra for different DC voltages from 0.5 to 7.5V are shown in Figure 3.

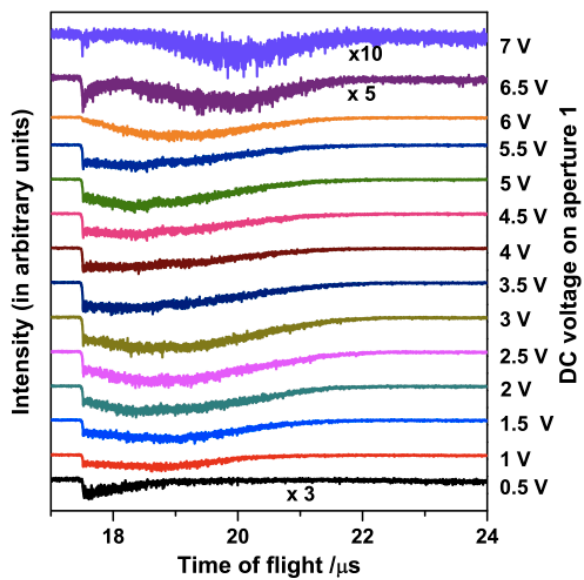

Figure 3: Time-of-flight spectra for different DC voltages applied to aperture 1 and 4.8 kV to ap. 2.

The question is what species are between aps. 1 and 2? Single pDFB cations passing the 4 mm slit will be immediately accelerated by the DC field. For a DC voltage of 0.5 V on ap.1, a pDFB$^+$ ion will only travel 22 mm before hitting the solid part of aperture 2, i.e. applying a voltage of 0.5 V will prevent any single ion to be detected. Individual very long-living ZEKE Rydberg states could, in principle, survive the flight time of ca. 500 µs from the nozzle to the TOF mass spectrometer. In the small DC electric field F such Rydberg states will be instantaneously ionized [16] according to the lowering of the ionization energy by 6 $\sqrt{F}$ [cm$^{-1}$] and the ions formed then behave in the same way as single cations, also hitting ap. 2. Besides individual Rydberg states also an ensemble of high-n and low-n Rydberg states could be produced. However, only the high-n Rydberg states could have a long enough lifetime to survive the flight time of 0.5 ms. A much better explanation is that this ensemble forms a plasma albeit one with a surprisingly long lifetime of > 0.5 ms. In the weak DC electric field this plasma starts decaying, releasing single ions from its outer parts while the central part continues moving (nearly) unaffected by the small field. Part of the signal observed in the TOF spectra thus originates from single cations released from the plasma, which still continues to travel parallel to aps. 1 and 2.

In the following we are discussing the observed time-of-flight spectra in Figure 3 in a more detailed way. The longest TOF's come from those ions between aps. 2 and 3 which are furthermost away from ap. 2 (closest to ap. 3) at the time when the HV pulse on ap. 2 is switched on. These ions represent the first ones released from the plasma that can be detected. The ions with the shortest TOF are those, which have just passed the grid of aperture 2 to enter the region between aps. 2 and 3 when pulse 2 is switched on. The observed step at 17.5 µs in the TOF spectra indicates that there are still ions emitted from the plasma but these cannot be detected, because once pulse 2 is switched on, these are reflected from ap.2 and prevented from entering the second acceleration stage between aps. 2 and 3. At a DC voltage of 0.5 V the peak maximum is located at 17.5 µs. With increasing voltage the maximum is shifted to longer flight times and seems to develop a double maximum structure. Interestingly, the step at 17.5 µs suddenly vanishes at a voltage of 6V and reappears at 6.5 V. At 6.5 V and 7 V clearly two separated peaks can be observed in the spectra. A definite explanation of the observed behavior is not possible due to the complexity of the systems and requires a lot of additional experimental and theoretical work. But one possible explanation for the observation, especially the multiple structures at 6.5 and 7 V, is that the plasma might have a shell-structure, with a very high density plasma core responsible for the signal at short flight times at 6.5 V and 7 V. This has been predicted in previous theoretical works [2]. We are presently performing simulations using SIMION 8 to investigate if the measured TOF spectra can be explained by a shell structure of the plasma.

So far, no plasma has been observed that after an unconfined expansion of 100 $\mu s$ has not lost its plasma characteristics [5]. In contrast, the plasma produced with the new method presented here reveals plasma properties even after *ca.* 0.5 *ms*. Together with the estimation of the Coulomb coupling parameter this indicates that the plasma is not only ultra-cold but also strongly-coupled [5].

In an additional experiment the influence of the position of the laser excitation related to the distance $z$ from the nozzle was investigated. For different $z$ the DC voltage was scanned, while the resulting ion current, integrated from 17.5 to 17.7$\mu s$, was measured. The results are displayed in Figure 4 for different excitation distances from the nozzle. For the distances closest to the nozzle two maxima are observed in the spectra, whereas at $z$=15 *mm* only one peak can be seen. At this larger distance the plasma decay is already observed at very low voltages/fields. At this distance the density of the $p$DFB$^+$ ions has dropped to $3.9*10^{12}$ $cm^{-1}$, resulting in an estimated ion Coulomb coupling parameter of ~60-210.

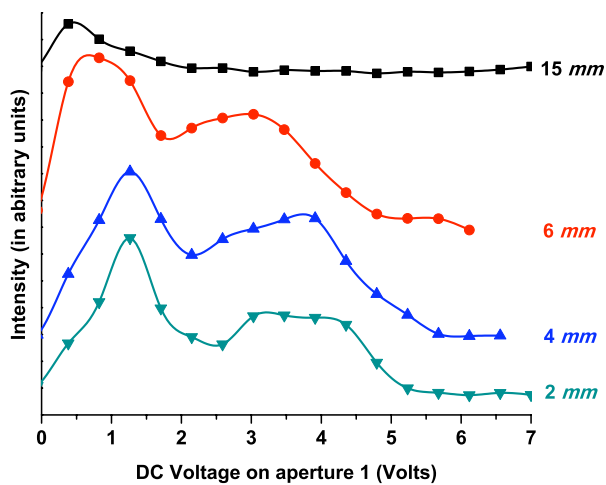

Figure 4: TOF signal integrated from 17.5 to 17.7$\mu s$ versus DC voltage on aperture 1 for 4.8 *kV* on grid aperture 2 for different distances from the jet nozzle.

There is another possible explanation for the multiple peaks in Figure 4. If the plasma, under the weak electric field, broke up into small pieces of partially charged plasma crystals or droplets with a different mass to charge ratio, which we tentatively name *"plaslets"*, these will move in the weak DC electric field, provided they also have a defined *centre of mass*. Once these partially charged *plaslets* are broken up into single cations by the strong electric field, this will then also contribute to the multiple peaks observed in Figure 4. Experimental results supporting this interpretation will be discussed further in a forthcoming paper [19].

In summary, we have shown that our method produces an ultra-cold plasma with a hitherto not observed lifetime of ~0.5 *ms*. The Coulomb coupling parameter was estimated to be around 230-820. The estimation of the electron temperature is much more complicated. We expect that in our plasma formation process the rate of collisions between electrons and ions is very low, leading to a very low TBR rate and therewith a low electron temperature. But at this point we cannot verify our assumption therefore the electron temperature is not discussed in this paper. The development of the electron temperature has to be further investigated and will be discussed in future publications. But the extreme long lifetime indicates that in this plasma not only the ions but also the electrons are strongly coupled. The experimental results may be interpreted in the way that the plasma has a shell structure and/or breaks up into smaller charged parts termed "plaslets". But these are only speculations at this time and will be investigate in a more detailed way.

It is a pleasure to thank E. Grant (Vancouver) and T. Killian (Houston) for discussions about the interpretation of our results. M. R. thanks The University of Manchester-Nuclear Decommissioning Authority Dalton Cumbria Project for support of this work.